\documentstyle[titlepage,12pt,cite]{article}
\newcommand{\be}{\begin{equation}}
\newcommand{\ee}{\end{equation}}
\newcommand{\bear}{\begin{eqnarray}}
\newcommand{\ear}{\end{eqnarray}}

\begin{document}
\begin{titlepage}
\rightline
{PITHA 97/3} \rightline{January 97} \rightline{ }

\vspace{0.8cm}
\begin{center}
{\bf\LARGE CP VIOLATION IN WEAK DECAYS AND ELSEWHERE \footnote{
Talk given at the 20th Johns Hopkins Workshop, Heidelberg, June 1996}\\}

\vspace{2cm}

\centerline{ \bf Werner Bernreuther},

\vspace{1cm}
\centerline{Institut f. Theoretische Physik,
RWTH Aachen, D-52056 Aachen, Germany}

\vspace{3cm}

{\bf Abstract:}\\
\parbox[t]{\textwidth}
{A brief overview is given on the status and prospects of searches 
for CP nonconservation
effects in weak decays of strange, charmed, and beauty hadrons, on the
search for permanent electric dipole moments
of particles, and on present and future high energy CP tests at colliders.
}

\end{center}
\vspace{2cm}

\end{titlepage}
\newpage

\section{Introduction}
The origin of CP nonconservation remains one of the few dark corners
of the 
theory of  electroweak
interactions. Another one is the dynamics of electroweak symmetry
breaking.
 Very probably
these two corners are related: clarification of weak gauge symmetry
breaking
 would also
shed light on the origin of CP violation. More pragmatically speaking,
we do
 not know so far whether this symmetry
 violation arises from a single 
``source" -- which is most likely the Kobayashi-Maskawa (KM) phase
\cite{KM} 
in the charged
weak quark currents -- or whether there are several CP-nonconserving 
interactions which will show
up in different physical situations.

There are a number of well-known reasons which motivate the belief
that 
the Standard Model (SM) is
part of a larger gauge theory.
Extensions of the SM almost invariably entail a larger non-gauge
sector -- 
i.e., scalar self interactions 
and Yukawa intercations -- than the SM.\footnote{For a recent review of CPV 
in the context of dynamical 
symmetry breaking, see
\cite{Inagaki}.}
In this  way quite a number of ``new" CP-violating (CPV) interactions 
\footnote{In the following, new CP-violating
interactions refer to interactions not due to the KM phase.}
for quarks  $and$ for leptons
are conceivable in a natural way. In particular CPV interactions with
the 
following features may exist:  
Interactions that are unrelated to the mixing of quark generations and
the 
hierarchy of quark masses. 
Well known
examples include CPV by an extended scalar Higgs 
sector \cite{Lee,Wein,DeMa,Bran,Wolf} and CPV phases in soft
supersymmetry 
(SUSY) 
breaking terms.   Such sources can induce also sizable CP effects in
flavour 
diagonal interactions.
Specifically, interactions involving Higgs bosons can induce effects
which
 drastically grow with some power of
a fermion mass, thus
leading to potentially large effects in the heavy flavour sector. So
far the 
only hint for CPV beyond 
the KM phase
are recent attempts to develop scenarios for explaining the baryon
asymmetry 
of the universe \cite{Cohen,FY}.

In the following a number of CPV laboratory phenomena due to the KM
phase and 
some non-KM sources of CP 
violation 
are discussed. (We shall not deal with the strong CP problem; for a
recent 
review, see \cite{Peccei}.)

\section{Weak Decays}
Observable CP violation \`a la KM requires quarks whose weak decays
are 
Cabibbo suppressed. That is not 
the case for $c$ and 
$t$ quarks. Therefore CP searches involving these quarks will
predominantly
 test for new interactions.

\subsection{Kaons and Hyperons}
The KM mechanism can account for the observed CP violation in $\Delta
S = 2$ 
$K^0 -\bar{K^0}$ mixing. 
The present experimental status on ``direct" $\Delta S = 1$ CP
violation in 
$K^0\to 2\pi$ is 
inconclusive \cite{eps1,eps2}. New 
experiments  aim at measuring $\rm{Re} (\epsilon'/\epsilon)$ at the
level of 
$10^{-4}$. On the 
theoretical side
considerable effort has ben spent over the last years to calculate the
next-to-leading order QCD corrections to the  effective weak
Hamiltonian 
within the SM, to pursue 
various approaches in 
determining weak matrix elements, and to get a handle on the various 
uncertainties involved in the 
prediction 
of $ \epsilon'/\epsilon$.
A recent detailed review \cite{Buras} of the current status estimates 
this  quantity within the SM  $\sim a\ few \times 10^{-4}$ .

Hyperon decays also offer a possibility to establish CP violation in 
$\Delta S = 1$ decays. Consider for instance the decay of polarized 
$\Lambda\to p\pi^-$ and ${\bar\Lambda}\to {\bar p}\pi^+$.
The differential decay distributions are proportional to 
$(1+\alpha_\Lambda\vec\omega_\Lambda\cdot\hat p_p)$
and $(1+\alpha_{\bar\Lambda}\vec\omega_{\bar\Lambda}\cdot\hat p_{\bar p})$, 
respectively, where $\vec\omega$ 
is the hyperon polarization vector and $\hat p$ is the (anti) proton 
direction of flight in the hyperon rest frame.
The spin analyser quality factor $\alpha$, which is parity-violating, is 
generated by the interference
of S and P wave amplitudes.  CP invariance requires
that $\alpha_\Lambda = - \alpha_{\bar\Lambda}$. Hence a CP observable is 
\begin{equation}
A_\Lambda = \frac{\alpha_\Lambda + \alpha_{\bar\Lambda}}
{\alpha_\Lambda - \alpha_{\bar\Lambda}}.
\label{Aslam}
\end{equation}
Note that $A_\Lambda$ is CP-odd but T-even, i.e., even under the
reversal 
of momenta and spins. Hence 
a non-zero asymmetry (\ref{Aslam}) requires, apart from CP phases,
also 
absorptive parts in the amplitudes.
Neglecting isospin $I=3/2$ contributions, an approximate expression
for 
$A_\Lambda$ 
is given by (see, for instance ref. \cite{Grimus}) 
\begin{equation}
A_\Lambda \simeq -
\tan(\delta^P_{1/2}-\delta^S_{1/2})\sin(\varphi^P_{1/2}-
\varphi^S_{1/2}),
\label{Aform}
\end{equation}
where $\delta^{S,P}_{1/2}$ and $\varphi^{S,P}_{1/2}$ are the S,P wave
final 
state phase shifts and weak
CP phases for the isospin $I=1/2$ amplitudes, respectively. 

In the Standard Model CP violation in $\Delta S = 1$ hyperon decays is
induced by penguin amplitudes. 
Extensions of
the SM may add charged Higgs penguin, gluino penguin contributions,
etc.  
Predictions for hyperon  
CP observables
like $A_\Lambda$ are usually obtained \cite{Don,HSV,HV} as follows:
within a 
given model of CP violation 
one computes 
first the effective weak $\Delta S = 1$ Hamiltonian at the quark level. 
(In the SM its next-to-leadig order QCD 
corrections are known \cite{Buras}.)  The strong phase
shifts $\delta^{S,P}_{1/2}$ are extracted from experimental data. The
usual 
strategy in 
determining the weak 
phases $\varphi^{S,P}_{1/2}$
is to take the real parts of the matrix elements 
$<\pi p|H_{eff}|\Lambda>^{S,P}_{I=1/2}$ 
from experiment, whereas 
the CPV part is computed using various models for hadronic matrix
elements. 
Although the 
theoretical uncertainties
 are quite large one may conclude \cite{HSV,HV} from these
calculations that 
within the 
SM the asymmetry
 $A_{\Lambda}$ is about $4\times 10^{-5}$.
Contributions from non SM sources of CP violation can yield larger
effects, 
but are  
constrained by the 
$\epsilon'$ and $\epsilon$ parameters
from $K$ decays. He and Valencia conclude that $|A^{non-SM}_{\Lambda}|$ cannot 
exceed $a\ few \times 10^{-4}$.

A high statistics hyperon  experiment \cite{Lambda}  (E871) at
Fermilab is 
underway. The decay
chain $\Xi^-\to\Lambda\pi^-\to p\pi^+\pi^-$ and the corresponding
decay 
chain for $\bar\Xi^+$ will be used.
They  $\Xi$'s will be produced  $unpolarized$. Then the $\Lambda$ 
polarization is given by 
${\vec\omega}_{\Lambda}$=$\alpha_{\Xi}
{ \hat p}_\Lambda$, where ${ \hat p}_\Lambda$ is the $\Lambda$
direction of 
flight in the $\Xi$ rest frame.
E871 measures the asymmetry 
\begin{equation}
A = \frac{\alpha_\Lambda\alpha_\Xi - \alpha_{\bar\Lambda}\alpha_{\bar\Xi}}
{\alpha_\Lambda\alpha_\Xi + \alpha_{\bar\Lambda}\alpha_{\bar\Xi}} 
\simeq A_\Lambda + A_\Xi.
\label{AslaX}
\end{equation}
$A_\Xi$ is estimated to be smaller than $A_\Lambda$ because of smaller
phase 
shifts. E871 
expect to produce 
about $10^9$ events. They aim at a sensitivity $\delta A \simeq
10^{-4}$. If 
an effect will 
be observed at this 
level it will be, in view of the above, most probably of non SM origin. 
 
\subsection{Charm}
$D^0 - \bar D^0$ mixing and associated CP violation in the $\Delta C =
2$ 
mixing amplitude,
and direct CP violation in the  $\Delta C = 1$ charm decay amplitudes
are 
predicted to be 
very small in the SM. 

In the SM direct CPV may be significant only for singly Cabibbo
suppressed 
decays. 
In this case one has at the 
quark level two contributions to the decay ampitude, namely the usual
``tree" 
amplitude and the 
penguin amplitude, that  have different
weak phases. At  the hadron level the decay  amplitude is of the form 
$A e^{i\delta_A} + B e^{i\delta_B}$, 
where $\delta_{A,B}$ are strong interaction phase shifts. This leads
to a 
CP asymmetry

\begin{equation}
A_D = \frac{\Gamma(D\to f) - \Gamma({\bar D}\to{\bar f})}
{\Gamma(D\to f) + \Gamma({\bar D}\to{\bar f})} \propto {\rm
Im}(AB^*)\sin 
(\delta_B - \delta_A).
\label{Asa}
\end{equation} 

Buccella et al. \cite{Buc} have investigated $A_D$ within the SM for a
 number of Cabibbo suppressed channels.
They calculated the strong phase phifts for the respective channels
by 
assuming dominance of the nearest 
resonance. For some modes, for instance $D^+\to {\bar K^{*0}}K^+$ and
$D^+\to 
\rho^+\pi^0$ they find
$A_D \sim 10^{-3}$. In some extensions of the SM like non-minimal 
supersymmetry \cite{Bigi} or 
left-right-symmetric models \cite{LeY} $A_D$ can be larger by about
one 
order of magnitude. 
Moreover,  asymmetries
of the same order could also be generated in these models for Cabibbo 
allowed and doubly Cabibbo suppressed channels. 

$D^0 - \bar D^0$ mixing  is very small in the SM, $x=\Delta
m_D/\Gamma_D
 << 10^{-2}$. However,
 quite a number of extensions of the SM, for instance multi-Higgs or 
supersymmetric extensions,   
can lead to $x \sim 10^{-2}$. In these models it is quite natural that
 there is (new) 
CP violation associated with 
$\Delta C = 2$ mixing. 
It is mostly these expectations \cite{Nir} from   SM extensions that 
nourish the hope
 of observable mixing and observable indirect and direct CP violation in 
 proposed high statistics charm 
experiments \cite{Kaplan} with $10^8$ to $10^9$ events.

\subsection{Beauty}

High statistics experiments with the aim of measuring CPV rate
asymmetries 
\cite{BG} in 
$B$ decays will provide, in a few
years,  the decisive test of the KM mechanism. These asymmetries are 
characterized by the angles --
conventionally called $\alpha, \beta,$ and $\gamma$ --  of the
well-known 
CKM unitarity triangle. 
Several fits \cite{Ali,Pich}, using input from CPV in $K$ decays, 
$B^0_d -{\bar B^0_d}$ 
mixing, etc., have been 
performed to constrain these angles. These fits yield in particular
$0.2 \le 
\sin(2\beta) \le 0.9$, 
supporting the expectation that CP violation outside the $K$ system
will 
first be observed through
an asymmetry between the rates of $B^0_d$ and ${\bar B^0_d} \to J/\Psi + K_S$. 
The integrated rate asymmetry, which can be calculated in a clean way,
is proportional to $\sin(2\beta).$

Similarly the time integrated rate asymmetry of $B^0_d, 
{\bar B^0_d} \to \pi^+ \pi^-$ is 
related to $\sin(2\alpha)$.
However, apart from the fact hat these modes have very small branching
ratios, there is an 
uncertainty in the prediction
of the CP asymmetry because of penguin diagrams contributing to the 
decay amplitudes. 
In principle this uncertainty 
can be eliminated by an isospin analysis \cite{GL}. (Recall that there
is no QCD 
penguin contribution to the $I = 3/2$ component
of the $B_d\to \pi\pi$ amplitude.) The  method requires measuring 
$B^0_d \to \pi^+ \pi^-, \pi^0 \pi^0$ and the
conjugated channels, 
and $B^\pm \to \pi^\pm \pi^0$.  It will be difficult to carry out.

The CP parameter $\sin(2\gamma)$ is for instance related to the time 
integrated asymmetry of
the rates  $B^0_s, {\bar B^0_s} \to \rho K_S$. However, that is not a
clean and feasible way  of
extracting $\sin(2\gamma)$: firstly because these
modes have very small branching ratios and secondly  because of 
theoretical complications in view of penguin contributions. One
proposed 
alternative is as follows \cite{GW}:
From the measured decay rates one has to determine the moduli of the
decay amplitudes for
$B^+\to D^0 K^+, {\bar D^0} K^+$, $D_{1,2} K^+$ and for the charge
conjugated 
channels. ($D_{1,2}$ are the
the CP- even and odd eigenstates.) From the two triangle relations
relating 
the three 
complex amplitudes for $B^+$ and for $B^-$, 
respectively,
one can obtain $\sin^2\gamma$ up to an ambiguity which can in
principle also 
be resolved. 

According to the  KM mechanism for the three generation SM 
 $\alpha + \beta + \gamma = \pi$. 
A deviation from this relation would provide evidence for new
CP-violating 
interactions \cite{BG2}. 
(If the sum of these angles turns out to be $\pi$, note that this does
not necessarily imply absence of
new CPV effects in the $B$ system.)
Of course, more specific searches for new CPV in the $B$ system can 
be made, for instance  by investigating CP observables that
are predicted to be small in the SM, e.g., the  asymmetry  in the rate for 
$B^0_s \to J/\Psi + \phi$ and its conjugated mode. 

\section{Electric Dipole Moments}

The searches for permanent electric dipole moments (EDM), for instance
of the neutron
or of an atom with non-degenerate ground are known to be   a very
sensitive  means to trace 
new CPV interactions.
Recall that a non-zero  EDM of a non-degenerate stationary state would
signal P and T violation; that is,
CP violation assuming  CPT invariance.

A non-zero atomic EDM $d_A$ could be due to a non-zero electron EDM
$d_e$,  non-zero nucleon EDMs, P- and
T-violating nucleon-nucleon, and/or electron-nucleon interactions.
Schematically,

\begin{equation}
d_A  = R_A d_e + C^{eN}_A +C^N_A.
\label{dat}
\end{equation} 
 It has been shown long ago \cite{Sandars} that paramagnetic atoms can
have large enhancement factors $R_A$.
More recent atomic physics calculations \cite{Liu} 
 obtained for instance for Thallium the factor $R_{Tl}\simeq -585$
with an estimated error of 
about 10$\%$.
For Thallium one has to good approximation $d_{Tl} 
\simeq d_e R_{Tl} + C^{eN}_{Tl}$. 
The nuclear contributions
can be neglected for the following reasons: The nuclear 
ground state of $^{205}{\rm Tl}$ has spin 1/2 and therefore cannot
have a nuclear 
quadrupole moment. A potential (small) contribution of
a Schiff moment of the Thallium nucleus is irrelevant at the present 
level of experimental sensitivity. 
From the experimental upper bound \cite{Commins} on $d_{Tl}$ and with 
$R_{Tl}$ the upper bound
$|d_e| < 4\cdot 10^{-27} e$ cm was derived \cite{Commins}. \\
Very precise experimental upper bounds were obtained on the EDMs of 
certain diamagnetic atoms, in
particular for mercury \cite{Hg}. The mecury EDM, like that of other 
diamagnetic
atoms, is not sensitive to $d_e$ but to the Schiff moment
of the $^{199}{\rm Hg}$ nucleus which at the quark-parton level would 
be due to non-zero (chromo) EDMs of quarks
and/or P- and T-violating quark-quark or gluonic effective
interactions. 
As the transition from the level
of partons to the level of a nucleus involves large uncertainties the 
experimental 
limits on the EDMs 
of diamagnetic atoms are difficult to interpret in terms of
microscopic 
models of 
CP violation \cite{Kat}.\\
Experimental searches for a non-zero EDM of the neutron  at 
Grenoble \cite{Gren} and 
at Gatchina \cite{Len} 
have lead to the upper limit $|d_n|<9\cdot 10^{-26} e$ cm. 

Theoretical predictions of the EDM of the electron -- or of other
leptons -- 
usually constitutes
a straightforward problem of perturbation theory because models of CPV
are  
weak coupling theories a posteriori.
However, a firm numerical prediction within a given extension of the
SM would  require knowledge of
parameters like masses and couplings of new particles, apart from CP phases. 
The calculation of $d_n$ and of T-violating nucleon-nucleon
interactions, 
etc. involves in addition
methodological uncertainties. For a given model of CPV one can usually
construct with reasonable precision
the relevant effective P- and T-violating low energy Hamiltonian at
the quark 
gluon level which contains
(chromo) EDMs of quarks, the $G\tilde G$ and $GG\tilde G$ terms,
etc. The 
transition to the nucleon/nuclear level,
that is, the computation of T-violating hadronic matrix elements
involves large uncertainties. In computing/estimating the neutron EDM
naive 
dimensional estimates,
the  quark and the MIT bag model \cite{McKellar}, sum rule 
techniques \cite{Chem,Khri1,Khri2}, 
and experimental constraints
on the quark contribution to the nucleon spin \cite{Flores} have been
used 
in particular.

The KM phase induces only tiny CP-violating effects in flavour-diagonal 
amplitudes. Hence the SM predicts tiny particle EDMs (barring the
strong CP 
problem of QCD; i.e., assuming
$\Theta_{QCD}= 0$).  A typical
estimate \cite{McKellar} for the neutron is $|(d_n)^{KM}| < 10^{-30}
e$ cm. 
In the SM with massless
neutrinos CPV in the lepton sector occurs only as a spill-over from
the quark sector: one 
estimates \cite{BeSu} that $|(d_e)^{KM}| < 10^{-37} e$ cm.

Quite a number of other CPV interactions are conceivable that lead to
neutron 
and electron  EDMs
of the same order of magnitude as the present experimental upper bounds. 
(For reviews, see \cite{McKellar,BeSu,Barr}.)  Multi Higgs extensions
of the 
SM can contain
neutral Higgs particles with indefinite CP parity. Exchange of these
bosons 
induces quark and lepton
EDMs already at one loop. For light quarks and leptons the dominant
effect 
occurs at two loops \cite{BaZe}.
In two-Higgs doublet extensions \cite{twol,Hay} of the SM with maximal
CPV 
in the neutral 
Higgs sector and a light
neutral Higgs particle with mass of order 100 GeV neutron and electron
EDMs 
as large as 
$10^{-25} e$ cm and
 $a\ few$ $\times 10^{-27} e$ cm, respectively, can be induced.

In the minimal supersymmetric extension of the SM (MSSM) there are in 
general, apart from the
KM phase, extra CP phases due to complex soft SUSY breaking
terms. These 
phases are  not bound to 
be small a priori. They generate quark and lepton EDMs and 
chromo EDMS of quarks at one-loop order \cite{EllFN,Flores,Barb} which
can  be quite large. (Unless the gaugino,
squark or slepton masses are close \cite{Kiz} to 1 TeV  which causes, 
however, other problems.) 
In particular, the prediction for the electron, which is not clouded by
hadronic uncertainties, is  $d_e \simeq 10^{-25} \sin{\varphi_e}\ (e$
cm) 
for neutralino and
$\tilde e$ masses of the order of 100 GeV. That means the leptonic
SUSY 
phase $\varphi_e$ must 
be quite small, which seems unnatural in the generic MSSM case. 
(For constrained versions see
for instance \cite{Cott}.)

In supersymmetric grand unified theories the small phase problem eases
by 
construction. 
In the SO(10) model considered in
refs. \cite{Hall,Barb} the phases in the soft terms are assumed to be
zero 
at the Planck
scale. Unification of the quarks and leptons of a generation into a
single 
multiplet leads, apart
from the KM phase, to extra CKM phases entering the fermion-sfermion 
gaugino (higgsino) 
interactions at the weak scale. 
GIM cancellations lead to a smaller $d_n$ and $d_e$ 
than in the generic MSSM -- but $d_e$ can be close to its experimental
upper bound.

Clearly, the present experimental EDM bounds have an impact on the 
parameter spaces of popular
extensions of the SM. In particular the bound on $d_e$ is important in
view of the
``theoretically clean" predictions. 
Further improvement of experimental sensitivity is highly desirable. 
As to future  low-energy T violation experiments:
A number of proposals \cite{Hinds,Weis} have been made to improve the 
experimental sensitivity 
to $d_e$ and to the EDMs of certain
atoms by factors of 10 to 100. There is also a new idea \cite{Golub} 
to measure the
neutron EDM with substantially improved sensitivity.

The present experimental sensitivity to EDMs of quarks and leptons
from 
the second and third fermion
generation is typically of the order $10^{-16}$ to $10^{-18} e$ cm (see below).
 Although this is orders of magnitude larger than the present limit 
on $d_e$ it constitutes 
nevertheless interesting
information. Some CP-violating interactions, for instance CPV Higgs
boson 
or leptoquark exchange, 
lead to EDMs in the heavy flavour sector that are much larger than 
$d_e$ or $d_n$.

\section{High Energy Searches}

Many proposals and studies for CP symmetry tests in high energetic 
$e^+e^-$, $p\bar p$,
and $p p$ colllisions have been made 
(see, for instance \cite{DV,B1,Gav,Stod,BN} for some early
studies). In particular the production and decay of $\tau$ leptons, 
$b$, and $t$ quarks 
are suitable for this purpose, as it allows for searches of new CPV 
interactions that become
stronger in the heavy flavour sector. Contributions from the KM phase
to the phenomena discussed below are negligibly small. Typically one 
pursues statistical tests with 
suitable asymmetries or correlations. One considers, for some
reaction, 
observables 
$\cal O_{CP}$ which change sign 
under a CP transformation. If the scattering amplitude of the reaction
is affected by
CPV interactions in a significant way then the interference of the 
CP-invariant and the CPV part
of the amplitude generates  non-zero expectation values $<{\cal O}_{CP}>$. 
Because an unpolarized $f\bar f$ state is a CP eigenstate in its
c.m. frame 
it can be shown \cite{BeNa}
that unpolarized (and transversely polarized) $e^+e^-$ and $p\bar p$ 
collisions allow for
``theoreticaly clean" CP symmetry tests: in these cases $<{\cal
O}_{CP}>$
 cannot be faked by CP-invariant
interactions as long as the phase space cuts are CP-blind. In the case of 
$p p$ collisions
potential contributions from CP-invariant interactions  to an
observable 
being used for a CP symmetry 
test (e.g., contributions from QCD absorptive parts to T-odd
quantities) 
must be carefully discussed.

In order to maximize the sensitivity to CPV couplings
it is often useful to consider so-called optimal observables \cite{AS}
that 
maximize the signal-to-noise ratio.
For a given reaction and a given model of CPV -- or a model
independent 
description of CPV using
effective Lagrangians or form factors -- with only one or a few small 
parameters these observables
can be constructed in a straightforward fashion.

\subsection{$e^+e^-\to\tau^+ \tau^-$}

CPV effects in tau lepton production with $e^+e^-$ collisions
 and in $\tau$ decay were discussed
in  \cite{Stod,BN,BBNO,GNelson,Nelson,BGV,AnRi,BBO,Choi,KuM}. 
CPV in 
$e^+e^-\to\tau^+ \tau^-$ can be traced back to non-zero EDM and weak dipole
moment (WDM) form factors \cite{BN,BBNO} $d^{\gamma}_{\tau}(s)$ and 
$d^{Z}_{\tau}(s)$, respectively, where $s = E^2_{c.m.}$. These form factors
induce a number of CP-odd tau polarization asymmetries and spin-spin 
correlations,
for instance a non-zero $d^{Z}_{\tau}(s)$ (more precisely, the real
part 
of that form factor) 
leads to a difference in the polarizations of $\tau^+$ and $\tau^-$ orthogonal 
to the scattering plane. Because the taus auto-analyse their spins
through 
their parity-violating
weak decays the tau polarization asymmetries and spin-spin
correlations 
transcribe to a number of CP-odd
angular correlations $<{\cal O}_{CP}>$ among the final states from
$\tau^+ 
\tau^-$ decay.

In their pioneering work the OPAL and ALEPH collaborations
\cite{O1,O2,A1,A2} 
at LEP have 
demonstrated that  CP tests in high energy $e^+e^-$ collisions can be 
performed with an accuracy at 
the few per mill
level. In the meantime the four LEP experiments measured a number of 
CP-odd correlations in
$e^+e^-\to\tau^+ \tau^-$. They turned out
to be consistent with zero. From these results upper limits on the
real
 and imaginary parts of the 
WDM form factors were derived. The combined upper limit
on the real part is \cite{Wermes} $|{\rm Re}d^{Z}_{\tau}(s=m^2_Z)| < 
3.6\cdot10^{-18}e$ cm (95$\%$ CL).

As already mentioned above the tau EDM and WDM form factors can be
much 
larger than the electron EDM.
There are a number of SM extensions where the dominant contributions
to 
these form factors are
one-loop effects, being not suppressed by small fermion masses. In these models
one has $d_\tau  = e\ \delta/m_Z$  whith $\delta$ of order $\alpha/\pi$.
For multi Higgs models one finds \cite{BBO} that $d_{\tau}$ 
can reach $10^{-20}e$ cm, whereas 
CPV scalar leptoquark exchange \cite{BBO} can lead to $d_{\tau}$ as
large 
as $3\cdot 10^{-19}e$ cm.

\subsection{$e^+e^-\to b\ {\bar b}\ gluon(s)$}
CP violation in this neutral current reaction would signal 
new interactions. At the parton level these interactions would affect 
correlations among
parton momenta/energies and parton spins. While the partonic momentum 
directions can be 
reconstructed from the  jet directions of flight the spin-polarization
of the $b$ quark
cannot, in general, be determined with reliable precision due to
fragmentation. This implies
that useful CP observables are primarily those which originate from
partonic 
momentum
correlations \cite{B1}. With these correlations only
chirality-conserving 
effective couplings
can be probed with reasonable sensitivity. Several correlations were 
proposed and 
studied \cite{B1,KO,B3,AL}. This situation is in contrast to
$\tau^+\tau^-$ 
and $t\bar t$ production
(see below) where the fermion polarizations can be traced in the
decays. 
That is why in these cases
searches for CPV dipole form factors, which are chirality-flipping,
can be 
made with good
precision.

In the framework of ${\rm SU}(2)_L$-invariant effective Lagrangians it
can 
be shown 
that chiral invariant CPV effective
$Zb{\bar b}G$ interactions of dimension $d=6$ (after spontaneous
symmetry
 breaking)
exist \cite{B1,BeNa}. In multi Higgs extensions of the SM these
interactions 
can be induced to
one-loop order \cite{BBHN}. They remain non-zero in the limit of 
vanishing $b$ quark mass.
Note that these CPV effective interactions are chiral-invariant $and$ 
flavour-diagonal which is a remarkable
feature. A dimensionless coupling $\hat h_b$ associated with these 
interactions \cite{B3}
turns out to be of the order of a typical one-loop radiative
correction, 
i.e., a few percent
if CP phases are maximal. This coupling could be larger in models with
excited quarks.

At the $Z$ resonance the above reaction provides an excellent
possibility to 
probe for
this type of interactions. The ALEPH collaboration \cite{Alep} has
recently 
made a CP 
study with their sample of $Z\to b{\bar b} G$ events. They obtained 
a limit of $|\hat h_b|<0.59$ at 95$\%$ CL.

\subsection{Top Quarks and Higgs Bosons}
Because of their extremely short lifetime top quarks decay on average before 
they can hadronize.
This means that the spin properties of $t$ quarks can be infered with good 
accuracy from their weak decays, i.e., $t\to W\ b$ in the SM.  
Like in the case of the tau lepton a number of $t$ spin-polarization
and spin-spin
correlation effects may be used to search for non-SM physics. Because of their
heavy mass top quarks, once they are available in sufficiently large numbers,
 will be a good probe of the electroweak symmetry breaking sector through their
Yukawa couplings. In particular they will be a good probe of Higgs
sector CP violation.
Many CP tests involving top quarks have been proposed.
These proposals include $t\bar t$ production in high energy $e^+e^-$ 
collisions \cite{BNOS,BSP,KLY,BO,ArSe,GRA,Rind,Pil1,Wien}
and in $p \bar p$ and $p p$ collisions 
\cite{BM,ScPe,BeBra2,Sch,Nacht,Atw,GradL}
at Tevatron and LHC energies, respectively. 
(As already mentioned,  in the latter case no genuine CP tests in the way 
desribed above can be made. One must carefully discuss and
compute potential fake effects.) Useful channels for these tests
are the final states from 
semileptonic decay of both $t$ and $\bar t$
and those from semileptonic   (nonleptonic)  $t ({\bar t})$ decay
 plus the charge conjugated channels.
(The charged lepton from semileptonic $t$ decay is known to be the most
efficient $t$ spin analyzer. Nonleptonic $t$ decays, on the other hand,
allow for reconstruction of the top momentum.) Observables ${\cal O}_{CP}$
include triple correlations, energy asymmetries, etc. 
and their  optimized versions. Computations of  $<{\cal O}_{CP}>$
have been made in a model-independent way using effective Lagrangians, 
form factor parameterizations
of the $t$ production and decay vertices, and within several extensions of
the SM, notably two-Higgs doublet and supersymmetric extensions. At the
upgraded Tevatron one can reach an interesting sensitivity
to the chromo EDM form factor of the top of about \cite{BM,Nacht,GradL}
$\delta d^{chromo}_t \simeq 10^{-18} e$ cm. Multi Higgs extensions of the SM
can  induce \cite{BSP,Soxu} top  EDM, WDM, and chromo EDM form factors
of this order
of magnitude. EDM and WDM form factors could be 
searched for most efficiently in $e^+e^-\to t\bar t$ not 
far above threshold \cite{BNOS,BO,Rind}. It was shown \cite{BO} 
that two-Higgs extensions of the SM induce CP effects at the percent level
in this reaction.

A possibility to check for CPV Yukawa couplings of the $t$ quark 
would be associated $t {\bar t}$ Higgs boson production. CP effects
can be large \cite{AS2} but the cross sections are quite small.

If neutral Higgs boson(s) $\varphi$ 
will be discovered and at least one of them 
can be  produced in reasonably large numbers then the CP properties
of the scalar sector could be determined directly by checking
whether $\varphi$ has $J^{PC}=0^{++},\ 0^{-+}$, or whether it
has undefined CP parity as predicted by multi Higgs extensions of the SM 
with Higgs sector CPV. A number of suggestions and theoretical
studies in this respect were 
made \cite{BBra1,Nor1,Chang,HMK,Kremer,DK,Seg,CoWi,GG,ABB,Pil}.
(Some of them follow the text book descriptions of how to determine
the CP parity of $\pi^0$.) In the fermion-antifermion decay of a
neutral Higgs particle with undefined CP parity CPV occurs at tree level
and manifests itself in a certain spin-spin correlation \cite{BBra1} which
can be as large as 0.5.
This correlation could be traced in $\varphi\to \tau^+\tau^-$ and,
for heavy $\varphi$, in $\varphi\to t \bar t$. In the latter case,
however, the narrow width approximation for $\varphi$ no longer
applies. Interference with 
the non-resonant background $t\bar t$ production \cite{BeBra2,BBra1}
diminishes the effect.

A ``Compton collider" realized by backscattering laser photons
off high energy $e^-$ or $e^+$ beams would be an excellent tool
to study Higgs bosons \cite{Zerwas} by tuning the beams to
resonantly produce $\varphi$. The CP properties of $\varphi$ could
be checked by appropriate asymmetries and correlations \cite{Kremer,GG,ABB}.

\section{Summary}

The gauge theory paradigm which describes  physics so well up to the
highest 
energy scales
presently attainable, and the circumstance that the electroweak
symmetry 
has to be broken spontaneously
allows, if there is physics beyond the Standard Theory, for a number
of different types of 
CP-violating interactions that would manifest themselves in different 
physical situations.
Hence searches for CP violation effects should be made in as many
particle 
reactions as possible.
$K$ decays and in the near future also hyperon decays may eventually 
establish direct
CP violation in weak transitions. In order to be able to discriminate 
better between models
improved calculations of hadronic matrix elements are needed.
The decisive tests of the KM mechanism will be provided by the $B$
meson factories
in a few years. The search for a neutron EDM, atomic EDMs, or other T-violation
effects in atoms or molecules remain a unique low energy window to
physics beyond the SM.
Searches of non-SM CP violation can also be made at present and future
high 
energy colliders.
In particular if Higgs sector CPV exists effects of up to  a few
percent are 
possible in the
top quark system. If Higgs boson(s) will be discovered and are
produced in 
large numbers 
it  is also conceivable to 
eventually study their CP properties directly. It is certainly an 
experimental  challenge to 
make  CP tests at the (sub) percent level at high energy hadron and 
future $e^+e^-$ colliders --
but it will be worthwhile to try.

\section*{Acknowledgments}
I wish to thank the organizers of the ``Johns Hopkins 96" workshop
for in\-viting me to this pleasant meeting.

\end{document}